\documentclass[11pt, oneside]{article} 
\pdfoutput=1
\usepackage{ytableau}
\ytableausetup{boxsize=0.8em}
\usepackage{jheppub}
\usepackage{amsmath,amssymb,amsfonts,amsthm}
\usepackage{color}
\usepackage{booktabs}
\definecolor{darkblue}{rgb}{0.1,0.1,.7}
\usepackage[]{amsmath}
\usepackage{graphicx} 

\usepackage[]{latexsym}
\usepackage{amscd}
\usepackage[all,cmtip]{xy}
\usepackage{mathrsfs}
\usepackage{bbold}
\usepackage[margin=10pt,font=small,labelfont=bf]{caption}
\usepackage{subcaption} 
\usepackage{simplewick}
\usepackage{changepage}
\usepackage[]{algorithm2e}
\usepackage{booktabs,multirow}
\usepackage{float}
 
\usepackage{hyperref}

\usepackage[OT2,OT1]{fontenc}
\usepackage{braket}
\usepackage{tikz}
\usepackage{pgfplots}
\pgfplotsset{compat=1.10}
\usepgfplotslibrary{fillbetween}
\usetikzlibrary{patterns}

\usepackage{tikz-feynman}
\NewDocumentCommand\semiloop{O{black}mmmO{}O{above}}
{%
	\draw[#1] let \p1 = ($(#3)-(#2)$) in (#3) arc (#4:({#4+180}):({0.5*veclen(\x1,\y1)})node[midway, #6] {#5};)
}

\usepackage[normalem]{ulem}

\newcommand{\abs}[1]{\left\lvert#1\right\rvert}

\theoremstyle{remark}

\def\XXint#1#2#3{{\setbox0=\hbox{$#1{#2#3}{\int}$}
     \vcenter{\hbox{$#2#3$}}\kern-.5\wd0}}

\makeatletter
\def\@fpheader{\ }
\makeatother

\title{Protected operators in non-local defect CFTs from AdS}
\author[a,b]{Jiaxin Qiao}
\affiliation[a]{Kavli Institute for the Physics and Mathematics of the Universe (WPI), \\ 
	The University of Tokyo Institutes for Advanced Study, The University of Tokyo, \\ 
	Kashiwa, Chiba 277-8583, Japan }
\affiliation[b]{Center for Data-Driven Discovery, Kavli IPMU (WPI), UTIAS, The University of Tokyo, Kashiwa, Chiba 277-8583, Japan}

\abstract{
	For a local quantum field theory in anti-de Sitter space with conformal boundary conditions but without dynamical gravity, the boundary theory is generically a non-local conformal field theory. Such theories can support conformal defects, but the standard local-CFT arguments based on a boundary stress tensor and conserved currents do not apply. We argue that, under general assumptions, displacement and tilt operators nevertheless exist and have protected quantum numbers. The mechanism is a Goldstone-type phenomenon in AdS: defect-induced symmetry breaking on the boundary is spontaneous from the viewpoint of the local bulk theory, whose Ward identities enforce the corresponding protected defect operators. We illustrate the mechanism in weakly coupled defect RG flows, long-range Landau--Ginzburg models, 4D Maxwell theory, and Yang--Mills theory in AdS.
}

\begin{document}
	
\maketitle

\section{Introduction}

The conformal bootstrap program, originating from the ideas of Polyakov \cite{Polyakov:1974gs}, has recently undergone a powerful revival, enabling precise, nonperturbative determinations of CFT data in $D \geq 3$ dimensions \cite{Rattazzi:2008pe,El-Showk:2012cjh,El-Showk:2014dwa,Poland:2018epd,Rychkov:2023wsd}. A natural extension of this framework arises in the presence of \emph{conformal defects} \cite{Affleck:1990by,McAvity:1995zd,DeWolfe:2001pq,Gliozzi:2015qsa,Billo:2016cpy}, which probe additional structures of the underlying bulk CFTs.

In recent years, increasing attention has been devoted to broader classes of
conformal theories that lie outside the standard paradigm of local CFTs. Notable
examples include the long-range Ising model and its generalizations
\cite{Paulos:2015jfa,Behan:2017emf,Chai:2021djc}, whose study dates back to
\cite{Fisher:1972zz,Sak:1973oqx,Sak1977}. These theories are \emph{non-local}
in the sense that their operator spectrum does not contain a conserved stress
tensor. This raises a basic question: to what extent do familiar structural
results of local CFTs survive in such non-local settings?

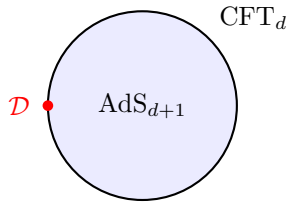
\begin{figure}
	\centering
	\begin{tikzpicture}[scale=0.5,
		>=Stealth,
		every node/.style={font=\small}
		]
		
		\fill[blue!8] (0,0) circle (2.5cm);
		
		\draw[thick] (0,0) circle (2.5cm);
		
		\node at (0, 0) {$\mathrm{AdS}_{d+1}$};
		
		\node[above right] at (45:2.5) {$\mathrm{CFT}_{d}$};
		
		\fill[red] (-2.5, 0) circle (4pt);
		
		\node[red, left=3pt] at (-2.5, 0) {$\mathcal{D}$};
		
	\end{tikzpicture}
	\caption{QFT in $\mathrm{AdS}_{d+1}$ with $\mathrm{CFT}_{d}$ as the
		conformal boundary condition, and the conformal defect $\mathcal{D}$
		is inserted on the boundary.}
		\label{fig:AdS_CFT_defect}
\end{figure}

In this work, we will be studying this question in the context of defect CFTs arising from Anti-de Sitter space (AdS), focusing on \emph{the existence of displacement and tilt operators}. The setup is as follows. Consider a local quantum field theory in AdS$_{d+1}$ with conformal boundary conditions, but without turning on the dynamical gravity. This defines a boundary theory that is typically non-local but still enjoys conformal symmetry \cite{Aharony:2015zea,Paulos:2016fap}. Such theories may support a conformal defect which is localized in a lower-dimensional submanifold (see figure~\ref{fig:AdS_CFT_defect}). This leads to a natural generalization of defect CFTs beyond the local framework.

The above defect-CFT setup is motivated by the studies of the confining flux tube in the Yang-Mills (YM) theory in AdS \cite{Gabai:2025hwf,Gabai:2026myo}. In this story, the bulk theory is YM with the magnetic boundary conditions \cite{Aharony:2012jf}, and the conformal defect is a large Wilson line inserted on the boundary. The Wilson line sources the flux tube in AdS, which has an effective string theory (EST) description at long distances \cite{Polchinski:1991ax}. The EST in AdS predicts Goldstone modes with the same quantum numbers as the displacement operator of a conformal line defect. This suggests that, from the YM perspective, the displacement operator should exist and have protected quantum numbers. However, since the EFT description is not first-principles, one still requires a mechanism within YM to explain this protection.

For a $p$-dimensional conformal defect in a $d$-dimensional \emph{local} CFT, the existence of displacement and tilt operators follows directly from the conservation of the stress tensor and the currents \cite{Jensen:2015swa,Bianchi:2018zpb}:
\begin{equation}\label{local_conservation}
	\begin{split}
		\partial^\mu T_{\mu i}(x)=-\delta^{(d-p)}(x_\perp)\mathbb{D}_i(x_{\parallel})\,,\quad \partial^\mu J_\mu^a(x)=-\delta^{(d-p)}(x_\perp)\mathbb{T}^a(x_\parallel)\,,
	\end{split}
\end{equation}
where $x_\parallel$ and $x_\perp$ denote the parallel and transverse directions of the conformal defect, $i$ is an index of the transverse direction and $a$ is an index of the global symmetry broken by the defect. Eq.\,\eqref{local_conservation} protects the scaling dimensions of displacement and tilt operators to be
\begin{equation}\label{dim_disp_tilt}
	\begin{split}
		\Delta_{\mathbb{D}}=p+1\,,\quad \Delta_\mathbb{T}=p\,.
	\end{split}
\end{equation}

Now back to the non-local CFT from AdS. The absence of a stress tensor and the currents invalidates \eqref{local_conservation}, so the above argument does not apply.

The main result of this work is that these operators nevertheless exist under
general assumptions, with their existence enforced by symmetry in the AdS bulk.
The key point is that, although the boundary theory need not contain a stress
tensor or conserved currents, the local bulk theory does, and the corresponding
Ward identities impose nontrivial constraints on boundary defects. The mechanism
may be viewed as a Goldstone phenomenon
\cite{Nambu:1960tm,Goldstone:1961eq,Goldstone:1962es} in AdS: a conformal
defect on the boundary breaks part of the boundary conformal symmetry, or of an
internal global symmetry, and this breaking is spontaneous in the bulk. The resulting Goldstone modes appear in the
defect spectrum as displacement and tilt operators, with protected quantum
numbers fixed by the symmetries.

This perspective applies to a wide range of systems including the above-mentioned long-range CFTs which are viewed as boundary conditions of free theories in AdS, as well as interacting bulk QFTs such as the $\phi^4$ theory or the gauge theories in AdS. In all these cases, the displacement and tilt operators emerge as protected operators.

The rest of the paper is organized as follows. In section~\ref{sec:argument}, we review the Ward–Takahashi identities relevant for spacetime and global symmetries, and present the argument for the existence of displacement and tilt operators as Goldstone modes associated with broken symmetries. In section~\ref{sec:example}, we discuss several examples, including weakly coupled models, Landau–Ginzburg-type theories, and gauge theories in AdS. We conclude in section~\ref{sec:conclusion} with a discussion of implications and open questions.

\paragraph{Note added.}
While finalizing this manuscript, the author became aware of closely related work by Lorenzo Bianchi, Elia De Sabbata and Marco Meineri \cite{Bianchi:2026}, who independently obtained the same results with similar arguments. Following mutual communication, we coordinated the submission so that the two papers appear on the same date.

\section{Goldstone modes in AdS}\label{sec:argument}
\subsection{Ward--Takahashi identities and spontaneous symmetry breaking}

Our starting point is the QFT correlators defined by the Euclidean path integral
\begin{equation}
	\begin{split}
		\braket{\mathcal{O}} &:= Z[g]^{-1}\int \mathcal{D}\phi\, \mathcal{O}[\phi,g]\, e^{-S[\phi,g]}\,, \\
		Z[g] &:= \int \mathcal{D}\phi\, e^{-S[\phi,g]}\,.
	\end{split}
\end{equation}
Here $\phi$ denotes the dynamical fields, $g$ is the background metric, and $\mathcal{O}$ is an observable constructed from $\phi$ and $g$. We will primarily consider observables given by products of local operators,
\[
\mathcal{O}=\mathcal{O}_1(x_1)\mathcal{O}_2(x_2)\cdots \mathcal{O}_n(x_n)\,.
\]
However, the discussion also applies to extended operators localized in a finite region, such as a Wilson loop of finite size.

We assume that the action $S[\phi,g]$ admits continuous global symmetries. The corresponding conservation laws follow from Noether's theorem,
\begin{equation}
	\nabla_\mu J_X^\mu = 0\,,
\end{equation}
where $J_X^\mu$ is the current associated with the symmetry generator $X$. At the level of correlation functions, this holds up to contact terms and leads to the Ward--Takahashi identities:
\begin{equation}\label{Wardid}
	\begin{split}
		\braket{\nabla_\mu J_X^{\mu}(x)\mathcal{O}}
		= -\sum_{i=1}^{n}\braket{\mathcal{O}_1(x_1)\cdots \left(Q_X\cdot\mathcal{O}_i\right)(x_i)\cdots\mathcal{O}_n(x_n)}\,\delta(x,x_i)\,,
	\end{split}
\end{equation}
where $Q_X$ denotes its action of $X$ on operators.

For spacetime symmetries, the currents take the form
\begin{equation}
	J_\xi^\mu = T^{\mu\nu}\,\xi_\nu\,,
\end{equation}
where $\xi$ is a Killing vector field and $T^{\mu\nu}$ is the stress tensor, satisfying $\nabla_\mu T^{\mu\nu}=0$ up to contact terms. The action of $\xi$ on operators is given by the Lie derivative,
\begin{equation}
	Q_{\xi}\cdot \mathcal{O}(x) = \mathcal{L}_\xi \mathcal{O}(x)\,.
\end{equation}

The identities \eqref{Wardid} are local. Integrating them over spacetime yields global symmetry relations for correlation functions. If no boundary contributions arise, one finds
\begin{equation}
	\sum_{i=1}^{n}\braket{\mathcal{O}_1(x_1)\cdots \left(Q_X\cdot\mathcal{O}_i\right)(x_i)\cdots\mathcal{O}_n(x_n)} = 0\,.
\end{equation}
In this case, the symmetry is preserved at the level of correlation functions. However, this relation can fail when boundary terms are nonvanishing, signaling spontaneous symmetry breaking (SSB).

A simple example of SSB is the massless free scalar, which has the shift symmetry $\phi \to \phi + \text{const}$ (equivalently, $Q\cdot \phi = 1$). The associated current is $J_\mu = \partial_\mu \phi$, and the Ward identity reads
\begin{equation}
	\braket{\partial_\mu J^\mu(x)\,\phi(y)} = -\delta(x-y)\,.
\end{equation}
Integrating over $x$, the boundary term on the left-hand side is nonzero and precisely reproduces the $-1$ on the right-hand side. This reflects spontaneous breaking of the shift symmetry, with the corresponding Goldstone mode identified with $\phi$ itself.\footnote{In flat space, spontaneous breaking of the shift symmetry occurs only in dimensions $D \geq 3$. In $D=2$, $\phi$ is not a well-defined operator in the infinite-volume limit, and the symmetry cannot be spontaneously broken, by the Mermin--Wagner--Hohenberg theorem \cite{Mermin:1966fe,Hohenberg:1967zz} and Coleman's ``no Goldstone" theorem \cite{Coleman:1973ci}.}

This illustrates a general mechanism for spontaneous symmetry breaking: nonvanishing boundary contributions encode the failure of charge conservation and lead to the emergence of Goldstone modes. In the AdS context, we will show that in the presence of the conformal defect on the boundary CFT, such breaking of spacetime and internal global symmetries gives rise to defect operators, identified respectively with the displacement and tilt operators.

\subsection{Assumptions and statements in AdS}\label{subsec:assumption_statement}

We work in Poincar\'e coordinates on AdS$_{d+1}$:
\begin{equation}\label{Poincare}
	\begin{split}
		ds^2 = R^2 \frac{dx^2 + dy^2 + dz^2}{z^2}\,, \qquad x \in \mathbb{R}^{p}\,, \quad y \in \mathbb{R}^{d-p}\,, \quad z \geq 0\,.
	\end{split}
\end{equation}
The boundary of AdS$_{d+1}$ is located at $z=0$. We decompose the boundary coordinates into $(x,y)$ for later convenience: in the presence of a defect, $y$ will parametrize the transverse directions.

\begin{figure}
	\centering
	\begin{subfigure}[b]{0.48\textwidth}
		\centering
		\begin{tikzpicture}[scale=1.3,
			>=Stealth,
			every node/.style={font=\normalsize}
			]
			
			\coordinate (O)  at (0.4,   0.2  );
			\coordinate (A)  at (0.4,   3.2);
			\coordinate (B) at (-1.6, 2.2);
			\coordinate (C) at (-1.6, -0.8);
			
			\fill[blue!12] (O) -- (A) -- (B) -- (C) -- cycle;
			\draw[thick]   (O) -- (A) -- (B) -- (C) -- cycle;
			
			\node[right] at (A) {$z=0$};
			\node             at (-0.3, 2.4)   {$\mathbb{R}^{d}$};

			\coordinate (Vpt) at (-0.8, 1.1);
			\fill (Vpt) circle (1.5pt);
			\node[above, xshift=2pt] at (Vpt) {$V(x,y)$};
			
			\coordinate (Ipt) at (1, 1.1);
			
			\fill (Ipt) circle (1.5pt);
			\draw[dashed, thick] (Vpt) -- (Ipt)
			node[right] {$\hat{\Phi}(z,x,y)$};
			
			\coordinate (Torigin) at (1.6, 2.4);
			\draw[->, thick] (Torigin) -- ++(0,   0.9) node[above] {$x$};
			\draw[->, thick] (Torigin) -- ++(0.55, 0.45) node[right] {$y$};
			\draw[->, thick] (Torigin) -- ++(1.0,  0  ) node[right] {$z$};
			
		\end{tikzpicture}
		\caption{Bulk-to-boundary expansion.}
		\label{fig:BOE}
	\end{subfigure}
	\hfill
	\begin{subfigure}[b]{0.48\textwidth}
		\centering
		\begin{tikzpicture}[scale=1.3,
			>=Stealth,
			every node/.style={font=\normalsize}
			]
			
			\coordinate (O)  at (0.4,   0.2  );
			\coordinate (A)  at (0.4,   3.2);
			\coordinate (B)  at (-1.6, 2.2);
			\coordinate (C)  at (-1.6, -0.8);
			
			\coordinate (Orig) at (-0.8, 1.1);
			
			\fill[blue!12] (O) -- (A) -- (B) -- (C) -- cycle;
			
			\draw[thick] (O) -- (A) -- (B) -- (C) -- cycle;
			
			\coordinate (Torigin) at (1.6, 2.4);
			\draw[->, thick] (Torigin) -- ++(0,   0.9) node[above] {$x$};
			\draw[->, thick] (Torigin) -- ++(0.55, 0.45) node[right] {$y$};
			\draw[->, thick] (Torigin) -- ++(1.0,  0  ) node[right] {$z$};
			
			\coordinate (Dbot) at (-0.8, -0.4);
			\coordinate (Dtop) at (-0.8,  2.6);
			\draw[red, very thick] (Dbot) -- (Dtop);
			
			\node[red, left]  at (-0.8, 2)  {$\mathcal{D}$};
			\node[right]      at (A)           {$z=0$};
			\node             at (-0.3, 2.4)   {$\mathbb{R}^{d}$};
			
			\coordinate (Vpt) at (-0.3, 1.1);
			\fill (Vpt) circle (1.5pt);
			\node[above, xshift=2pt] at (Vpt) {$V(x,y)$};
			
			\coordinate (Ipt) at (1, 1.1);
			
			\fill (Ipt) circle (1.5pt);
			\draw[dashed, thick] (Vpt) -- (Ipt)
			node[right] {$\hat{\Phi}(z,x,y)$};
			
			\coordinate (dpt) at (-0.8, 0.85);
			\fill (dpt) circle (1.5pt);
			\draw[dashed, thick] (Vpt) -- (dpt)
			node[left] {$\mathcal{O}(x)$};
			
			\draw[dashed, thick] (Ipt) -- (dpt);
			
		\end{tikzpicture}
		\caption{Bulk-to-boundary and bulk/boundary-to-defect.}
		\label{fig:conformal_defect}
	\end{subfigure}
	\caption{The operator expansions without (left) and with (right) a conformal defect.}
	\label{fig:combined}
\end{figure}
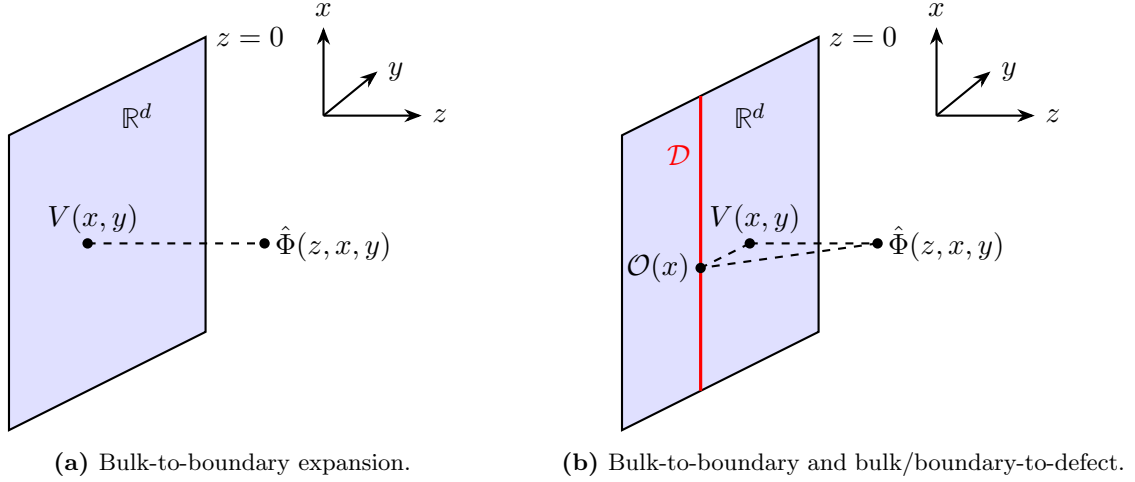

We consider a local QFT in AdS$_{d+1}$ with a conformal boundary condition, characterized by the following assumptions:

\begin{itemize}
	\item[(A1)] \textbf{AdS invariance.} Correlation functions are invariant under the AdS isometry group:
	\begin{equation}
		\begin{split}
			\braket{\hat{\Phi}_1(X_1)\hat{\Phi}_2(X_2)\ldots}
			=
			\braket{(g\cdot\hat{\Phi}_1)(X_1)(g\cdot\hat{\Phi}_2)(X_2)\ldots}\,,
			\qquad \forall g \in SO(1,d+1)\,.
		\end{split}
	\end{equation}
	This invariance follows from the Ward identities associated with the stress tensor in AdS, assuming that no boundary contributions arise upon integration.
	
	\item[(A2)] \textbf{Conformal boundary.} The boundary condition is described by a (generically non-local) CFT$_d$. Bulk operators admit an expansion in terms of boundary operators:
	\begin{equation}\label{eq:BOE}
		\begin{split}
			\hat{\Phi}(z,x,y)
			=
			\sum_{V} C_{\hat{\Phi}V}\, z^{\Delta_V - \Delta_{\hat{\Phi}}}\, V(x,y)\,,
		\end{split}
	\end{equation}
	where the sum runs over a basis of boundary operators $V$ (see figure~\ref{fig:BOE}). We choose a basis of bulk and boundary operators that diagonalizes dilatations:
	\begin{equation}\label{Op_scaling}
		\begin{split}
			(\lambda^{D}\cdot \hat{\Phi})(z,x,y)
			&= \lambda^{\Delta_{\hat{\Phi}}}\, \hat{\Phi}(\lambda z,\lambda x,\lambda y)\,, \\
			(\lambda^{D}\cdot V)(x,y)
			&= \lambda^{\Delta_V}\, V(\lambda x,\lambda y)\,.
		\end{split}
	\end{equation}
	
	\item[(A3)] \textbf{Internal global symmetry.} The correlation functions are invariant under a global symmetry group $G$, which commutes with the AdS isometry group,
	\[
	[G, SO(1,d+1)] = 0\,.
	\]
	This invariance follows from Ward identities associated with the current in AdS, assuming no boundary contributions upon integration.
\end{itemize}

We emphasize that the local stress tensor and conserved currents are defined in the bulk of AdS$_{d+1}$ and are generically absent on the boundary. More precisely, the boundary theory does not contain a spin-2 operator of dimension $d$ (the stress tensor), nor a spin-1 operator of dimension $d-1$ (the conserved current).

We now introduce a $p$-dimensional conformal defect $\mathcal{D}$, located at $z = 0$ and $y = 0$. By ``conformal defect" we mean that the defect preserves the spacetime subgroup symmetries
\begin{equation}
	\begin{split}
		SO(1,p+1) \times SO(d-p) \subset SO(1,d+1)\,,
	\end{split}
\end{equation}
where $SO(1,p+1)$ is the conformal group acting on $\mathbb{R}^p$ (extended to $\mathbb{R}^{d}$), and $SO(d-p)$ is the rotation group in the transverse space $\mathbb{R}^{d-p}$. If the original theory has an additional global symmetry $G$, the defect may preserve a subgroup $H \subseteq G$.

\begin{table}[t]
	\centering
	\begin{tabular}{c|c}
		\hline
		Local CFT with conformal defect & AdS with boundary conformal defect \\
		\hline
		Bulk--bulk OPE & Boundary--boundary OPE \\
		Bulk-to-defect expansion & Boundary-to-defect expansion \\
		Defect--defect OPE & Defect--defect OPE \\
		--- & Bulk-to-boundary expansion \\
		--- & Bulk-to-defect expansion \\
		\hline
	\end{tabular}
	\caption{Operator expansions in a local CFT and in a local QFT in AdS in the presence of a conformal defect. Boundary operators in AdS play the same role as bulk operators in the CFT.}
	\label{tab:expansions}
\end{table}

The operator expansions in the presence of a conformal defect are summarized in table~\ref{tab:expansions}. In a local CFT, these expansions are naturally understood from the cutting-and-gluing picture of the path integral \cite{Simmons-Duffin:2016gjk}. An analogous picture applies to a local QFT in AdS. In this framework, the boundary-to-defect expansion in AdS can be obtained from bulk-to-defect expansion by moving the bulk operator to the boundary (with proper rescaling).

Based on these considerations, we modify assumptions (A1)--(A3) in the presence of a conformal defect as follows:
\begin{itemize}
	\item[(A1')] Correlation functions are invariant under $SO(1,p+1)\times SO(d-p)$;
	\item[(A2')] Assumption (A2) continues to hold away from the defect, with the same coefficient functions. In addition, AdS bulk operators admit an expansion in terms of operators in the defect CFT$_p$ (see figure~\ref{fig:conformal_defect});
	\item[(A3')] Correlation functions are invariant under global symmetry $H \subseteq G$.
\end{itemize}

At this stage, we do not assume that the full conformal group $SO(1,d+1)$ is necessarily broken. In some special cases, it may remain preserved in the presence of the defect (for instance, for topological defects \cite{Verlinde:1988sn,Petkova:2000ip,Gaiotto:2014kfa}).

Under the above assumptions, our claim is that when the boundary conformal symmetry is explicitly broken to $SO(1,p+1)\times SO(d-p)$ by the defect, there must exist displacement operators $\mathbb{D}_i$ in the defect CFT, characterized by the following properties:
\begin{itemize}
	\item $\mathbb{D}_i$ is a scalar primary operator with scaling dimension $p+1$ and transverse spin $1$,\footnote{Transverse spin means the representation of the transverse rotation group $SO(d-p)$. In particular, transverse spin 0 and spin 1 mean scalar and vector representations.} where $i=1,\ldots,d-p$ labels the transverse directions;
	\item any infinitesimal global translation in the transverse directions can be implemented by inserting the displacement operator integrated over the defect:
	\begin{equation}\label{displacement_transformation}
		\begin{split}
			\int d^p x\, \braket{\mathbb{D}_i(x)\,\Phi}_{\mathcal{D}} + \braket{\mathcal{L}_{\xi_i}\Phi}_{\mathcal{D}} = 0\,, \qquad \forall \Phi\,.
		\end{split}
	\end{equation}
\end{itemize}

A similar statement applies when the defect breaks continuous internal global symmetries. If $H \subset G$ is the unbroken subgroup, then for each broken generator there exists a tilt operator $\mathbb{T}^a$ in the defect CFT with the following properties:
\begin{itemize}
	\item $\mathbb{T}^a$ is a scalar primary operator with scaling dimension $p$ and transverse spin $0$, where $a$ labels the broken generators of $G$;
	\item any infinitesimal transformation generated by a broken symmetry can be realized by inserting the tilt operator integrated over the defect:
	\begin{equation}\label{tilt_transformation}
		\begin{split}
			\int d^p x\, \braket{\mathbb{T}^a(x)\,\Phi}_{\mathcal{D}} + \braket{Q^a \Phi}_{\mathcal{D}} = 0\,, \qquad \forall \Phi\,.
		\end{split}
	\end{equation}
\end{itemize}

Before presenting the argument, let us emphasize the assumption that $G$ commutes with $SO(1,d+1)$. This condition is crucial for the conclusion $\Delta_\mathbb{T} = p$. If it is violated, the tilt operator need not have scaling dimension $p$. One explicit example is the shift symmetry of a scalar field in AdS, which we discuss in section~\ref{app:shift}.

We now turn to the derivation of these statements. For pedagogical clarity, we will present the argument in sections~\ref{subsec:displacement} and \ref{subsec:tilt}, with certain caveats on the convergence of the operator expansions. We leave an improved argument in section~\ref{app:improve}.

\subsection{Broken conformal symmetry $\Rightarrow$ displacement operator}\label{subsec:displacement}
We assume that conformal symmetry breaking happens on the boundary. Let us first argue that symmetry breaking on the boundary necessarily implies symmetry breaking in the bulk. This follows from assumption (A2'), that the bulk correlators admit an expansion in terms of boundary correlators, with the same structure as in the absence of the defect. Suppose, for contradiction, that the correlation functions in the bulk preserve the AdS isometry. Expanding them into power series of $z$ would then imply that the boundary correlators are also invariant under the full conformal group, contradicting our assumption.

Therefore, when the conformal defect breaks the full boundary conformal symmetry, there must exist bulk correlation functions that fail to respect the broken AdS isometries:
\begin{equation}\label{eq:brokenKilling}
	\begin{split}
		\sum_{i=1}^{n}\braket{\mathcal{O}_1(X_1)\cdots \mathcal{L}_\xi \mathcal{O}_i(X_i)\cdots \mathcal{O}_n(X_n)}_{\mathcal{D}} \neq 0\,,
	\end{split}
\end{equation}
where $\xi$ is a Killing vector corresponding to a broken AdS isometry.

We focus on the broken translation generators along the transverse directions of the defect. In Poincar\'e coordinates \eqref{Poincare}, the corresponding Killing vectors are
\begin{equation}
	\begin{split}
		\xi_i = \frac{\partial}{\partial y^i}\,, \qquad i=1,2,\ldots,d-p\,.
	\end{split}
\end{equation}

\begin{figure}
	\centering
	\begin{tikzpicture}[scale=1.0,
		>=Stealth,
		every node/.style={font=\normalsize}
		]
		
		
		\draw[thick] (0, -1.8) -- (0, 2.2);
		\node[above] at (0, 2.2) {$z\!=\!0$};
		
		\fill[red] (0, 0) circle (2pt);
		\node[red, left=2pt] at (0, 0) {$\mathcal{D}$};
		
		\draw[blue, thick] (2.3, 0.0) ellipse (1.8cm and 1.7cm);
		
		\node[blue, above] at (2.3, 1.6)
		{$\displaystyle\int T_{ni}\, d\Sigma_n$};
		
		\node[black] at (1.6,  0.9) {$\times\mathcal{O}_1$};
		\node[black] at (3.2,  0.8) {$\times$};
		\node[black] at (1.3, -0.2) {$\times\mathcal{O}_2$};
		\node[black] at (3.5, -0.3) {$\times\mathcal{O}_n$};
		\node[black] at (2.2, -0.9) {$\cdots$};
		\node[black] at (2.8, -1.0) {$\times$};
		
		\node[font=\Large] at (5.2, 0) {$=$};
		
		\def\rx{6.4}
		
		\draw[thick] (\rx, -1.8) -- (\rx, 2.2);
		\node[above] at (\rx, 2.2) {$z\!=\!0$};
		
		\fill[red] (\rx, 0) circle (2pt);
		\node[red, left=2pt] at (\rx, 0) {$\mathcal{D}$};
		
		\draw[blue, thick] (\rx+0.5, -1.8) -- (\rx+0.5, 2.2);
		
		\node[blue, above, align=center] at (\rx+2.3, 1.7)
		{$\displaystyle -\int T_{zi}\, \frac{d^p x\, d^{d-p}y}{z^{d-1}}$};
		
		\node[black] at (\rx+1.6,  0.9) {$\times\mathcal{O}_1$};
		\node[black] at (\rx+3.2,  0.8) {$\times$};
		\node[black] at (\rx+1.3, -0.2) {$\times\mathcal{O}_2$};
		\node[black] at (\rx+3.5, -0.3) {$\times\mathcal{O}_n$};
		\node[black] at (\rx+2.2, -0.9) {$\cdots$};
		\node[black] at (\rx+2.8, -1.0) {$\times$};
		
	\end{tikzpicture}
	\caption{Integrating the Ward--Takahashi identity: enclosing operators
		with a surface $\Sigma_n$ (left) equals the contribution from the
		constant-$z$ slice (right).}
	\label{fig:Tint}
\end{figure}
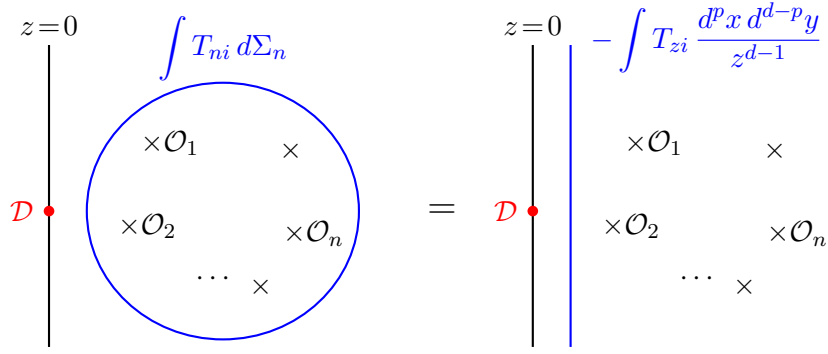

Let $\Phi = \mathcal{O}_1(x_1)\mathcal{O}_2(x_2)\cdots \mathcal{O}_n(x_n)$. Using the Ward--Takahashi identity \eqref{Wardid} with $J_{\xi_i}^\mu = T^{\mu}_{\ i}$, the violation of translation symmetry can be rewritten as
\begin{equation}\label{TO:integral}
	\begin{split}
		\int \frac{d^p x\, d^{d-p} y}{z^{d-1}} \, \braket{T_{z i}(z,x,y)\,\Phi}_{\mathcal{D}}
		=
		\braket{\mathcal{L}_{\xi_i}\Phi}_{\mathcal{D}}\,,
	\end{split}
\end{equation}
where $T_{zi}$ denotes the $z$-$y^i$ component of the bulk stress tensor. The radial coordinate $z$ is chosen to be smaller than all radial coordinates $z_i$ appearing in $\Phi$ (see figure~\ref{fig:Tint}). We assume that the contribution from the contour at infinity vanishes. The factor of $z^{d-1}$ in the measure arises from
\begin{equation}
	\begin{split}
		\sqrt{g}\, T^{z}{}_{i} = \sqrt{g}\, g^{zz} T_{zi} = \frac{1}{z^{d-1}}\, T_{zi}\,.
	\end{split}
\end{equation}
For the discussion here, we omit the indices of the internal global symmetries for convenience, since $T_{zi}$ transforms as a singlet of that.

We now make full use of assumption (A2'), which allows us to expand $T_{zi}$ in terms of defect operators:
\begin{equation}\label{T:bulk_to_defect}
	\begin{split}
		T_{zi}(z,x,y) = \sum_{\mathcal{O}} C^{T\mathcal{O}}_{zi,I}(z,y)\left[\mathcal{O}^I(x) + \ldots \right]\,.
	\end{split}
\end{equation}
Here the sum runs over multiplets of $SO(1,p+1)\times SO(d-p)$, whose primaries are scalars under $SO(p)$. The index $I$ labels a representation of $SO(d-p)$, and the ``$\ldots$'' denotes descendants under $SO(1,p+1)$. The coefficient function $C^{T\mathcal{O}}_{zi,I}(z,y)$ is constrained by the unbroken conformal symmetries $SO(1,p+1)\times SO(d-p)$. In particular, the transformation properties of $T$ and $\mathcal{O}$ under dilatations imply:
\begin{equation}
	\begin{split}
		&T_{zi}(z,x,y)\to \lambda^2 T_{zi}(\lambda z,\lambda x,\lambda y)\,, \qquad 
		\mathcal{O}^I(x)\to \lambda^{\Delta_{\mathcal{O}}}\mathcal{O}^I(\lambda x)\,, \\
		&\Longrightarrow \quad 
		C^{T\mathcal{O}}_{zi,I}(\lambda z,\lambda y)
		= \lambda^{\Delta_{\mathcal{O}} - 2} \, C^{T\mathcal{O}}_{zi,I}(z,y)\,.
	\end{split}
\end{equation}

Most defect operators in \eqref{T:bulk_to_defect} do not contribute to the integral in \eqref{TO:integral}. To see this, consider the contribution from a single multiplet:
\begin{equation}
	\begin{split}
		\int d^p x\, d^{d-p}y\, C^{T\mathcal{O}}_{zi,I}(z,y)\left[\mathcal{O}^I(x) + \alpha(z/\abs{y}) z^2 \partial_x^2 \mathcal{O}^I(x) + \ldots \right]\,.
	\end{split}
\end{equation}
We make the following observations:
\begin{itemize}
	\item Only defect primaries contribute, since descendants give total derivatives in $x$, which vanish upon integration.
	
	\item Only operators with transverse spin 1 (i.e. $\ydiagram{1}$ of $SO(d-p)$) contribute. For any other $SO(d-p)$ representation $\rho_{\mathcal{O}}$, the tensor product $\rho_{\mathcal{O}}\otimes \ydiagram{1}$ does not contain the trivial representation. Consequently, the coefficient function $C^{T\mathcal{O}}_{zi,I}(z,y)$, viewed as a function of the angular variables of $y$, decomposes into nontrivial spherical harmonics on $S^{d-p-1}$. Since these integrate to zero, we have
	\begin{equation}
		\begin{split}
			\int d^{d-p}y\, C^{T\mathcal{O}}_{zi,I}(z,y) = 0\,, \qquad \forall\, \rho_{\mathcal{O}} \neq \ydiagram{1}\,.
		\end{split}
	\end{equation}
	
	\item For operators with transverse spin 1, there are two independent $SO(d-p)$ tensor structures:
	\begin{equation}
		\begin{split}
			C^{T\mathcal{O}}_{\mu i,j}(z,y)
			=
			f^{T\mathcal{O}}_0\!\left(\tfrac{y^2}{z^2}\right) z^{\Delta_{\mathcal{O}}-2} \delta_{ij}
			+
			f^{T\mathcal{O}}_1\!\left(\tfrac{y^2}{z^2}\right) z^{\Delta_{\mathcal{O}}-2}
			\left(\frac{y_i y_j}{y^2} - \frac{\delta_{ij}}{d-p}\right)\,.
		\end{split}
	\end{equation}
	Only the first term survives upon integration over the angular directions of $y$.
\end{itemize}

Combining these observations, we obtain
\begin{equation}
	\begin{split}
		\int d^p x\, d^{d-p}y\, T_{zi}(z,x,y)
		=
		\sum_{\mathcal{O}\in\ydiagram{1}}
		\int d^p x\, d^{d-p}y\,
		f^{T\mathcal{O}}_0\!\left(\tfrac{y^2}{z^2}\right)
		z^{\Delta_{\mathcal{O}}-2}
		\mathcal{O}_i(x)\,.
	\end{split}
\end{equation}

We now perform the integration over the transverse directions. By dimensional analysis,
\begin{equation}\label{fTOint_scaling}
	\begin{split}
		\int d^{d-p}y\, f^{T\mathcal{O}}_0\!\left(\tfrac{y^2}{z^2}\right)
		=
		F^{T\mathcal{O}}\, z^{d-p}\,,
	\end{split}
\end{equation}
where $F^{T\mathcal{O}}$ is a constant. Substituting this into \eqref{TO:integral}, we find
\begin{equation}\label{TO_sum_rule}
	\begin{split}
		\sum_{\mathcal{O}\in\ydiagram{1}}
		F^{T\mathcal{O}}\, z^{\Delta_{\mathcal{O}}-p-1}
		\int d^p x\, \braket{\mathcal{O}_i(x)\Phi}_{\mathcal{D}}
		=
		\braket{\mathcal{L}_{\xi_i}\Phi}_{\mathcal{D}}\,.
	\end{split}
\end{equation}

This relation holds for any $0 < z < \min_i z_i$, where $z_i$ are the radial positions of the operators in $\Phi$. Since the right-hand side is nonzero and independent of $z$, it follows that:
\begin{itemize}
	\item at least one term on the left-hand side does not vanish;
	\item all nonvanishing contributions must satisfy $\Delta_{\mathcal{O}} = p+1$.
\end{itemize}

The above procedure uniquely isolates operators with the quantum numbers of the displacement operator, allowing us to define the defect operator
\begin{equation}\label{def:displacement}
	\begin{split}
		\mathbb{D}_i(x)
		:=
		-\sum_{\substack{\mathcal{O}\in\ydiagram{1} \\ \Delta_{\mathcal{O}}=p+1}}
		F^{T\mathcal{O}}\, \mathcal{O}_i(x)\,,
		\qquad i=1,\ldots,d-p\,.
	\end{split}
\end{equation}
By construction, $\mathbb{D}_i$ is a scalar primary of $SO(1,p+1)$ with scaling dimension $p+1$ and transverse spin 1. Moreover, the right-hand side of \eqref{def:displacement} is independent of the choice of bulk observable $\Phi$. Therefore, for any $\Phi$, the relation \eqref{displacement_transformation} holds, which is the defining property of the displacement operator associated with the conformal defect.

\subsection{Broken internal symmetry $\Rightarrow$ tilt operator}\label{subsec:tilt}

The argument for the existence of tilt operators closely parallels that for displacement operators.\footnote{The argument here can be viewed as a defect generalization of \cite{Porrati:2024zvi,Ankur:2026ylr}, where the case $p=d$ was discussed.} The main difference is that the stress tensor $T_{zi}$ is replaced by the conserved current $J^a_{z}$, which transforms as a scalar under $SO(p)\times SO(d-p)$ and exhibits a different scaling behavior under dilatations:
\begin{equation}
	\begin{split}
		J^a_{z}(z,x,y)\to \lambda\, J^a_{z}(\lambda z,\lambda x,\lambda y)\,.
	\end{split}
\end{equation}
As a result, the coefficient functions in the defect expansion of $J^a_{z}$ scale differently with $z$:
\begin{equation}
	\begin{split}
		J^a_{z}(z,x,y)
		=
		\sum_{\mathcal{O}\in \bullet \otimes \rho_J}
		g^{J\mathcal{O}}_0\!\left(\tfrac{y^2}{z^2}\right)
		z^{\Delta_{\mathcal{O}}-1}
		\left[\mathcal{O}^a(x)+\ldots\right]
		+\ldots\,,
	\end{split}
\end{equation}
where $\bullet$ denotes the scalar representation of $SO(1,p+1)\times SO(d-p)$, and $\rho_J$ denotes the representation of the unbroken subgroup $H$ under which the broken generators transform. The first ``$\ldots$'' denotes descendants under $SO(1,p+1)$, while the second ``$\ldots$'' denotes contributions from operators in other representations of $SO(1,p+1)\times SO(d-p)\times H$.

Repeating the analysis of the previous subsection, we obtain
\begin{equation}
	\begin{split}
		\sum_{\mathcal{O}\in \bullet \otimes \rho_J}
		G^{J\mathcal{O}}\, z^{\Delta_{\mathcal{O}}-p}
		\int d^p x\, \braket{\mathcal{O}^a(x)\Phi}
		=
		\braket{Q^a \Phi}\,,
	\end{split}
\end{equation}
where $G^{J\mathcal{O}}$ is defined by
\begin{equation}
	\begin{split}
		\int d^{d-p}y\, g^{J\mathcal{O}}_0\!\left(\tfrac{y^2}{z^2}\right)
		=
		z^{d-p} G^{J\mathcal{O}}\,.
	\end{split}
\end{equation}
From this relation, it follows that:
\begin{itemize}
	\item at least one term on the left-hand side does not vanish;
	\item all nonvanishing contributions satisfy $\Delta_{\mathcal{O}} = p$.
\end{itemize}

We therefore define the defect operator
\begin{equation}
	\begin{split}
		\mathbb{T}^a(x)
		:=
		-\sum_{\substack{\mathcal{O}\in \rho_J \\ \Delta_{\mathcal{O}}=p}}
		G^{J\mathcal{O}}\, \mathcal{O}^a(x)\,.
	\end{split}
\end{equation}
It is a scalar primary of $SO(1,p+1)\times SO(d-p)$ with scaling dimension $p$, transforming in the representation $\rho_J$ of $H$ (the representation carried by the broken generators). This operator satisfies \eqref{tilt_transformation}, the defining property of the tilt operator.

\subsection{An improved argument for Goldstone modes}\label{app:improve}

In sections~\ref{subsec:displacement} and \ref{subsec:tilt}, we argued that spontaneously broken symmetries imply the existence of protected operators, namely the displacement and tilt operators. The main idea was to use the assumed bulk-to-defect expansion and identify the Goldstone modes responsible for the symmetry breaking. However, this argument involves an important caveat: we implicitly assumed that the bulk-to-defect expansion converges everywhere, and that it is legitimate to interchange this expansion with the surface integral of the stress tensor or conserved current.

Experience from AdS/CFT suggests that such assumptions are not always satisfied. Even in the absence of a defect, the bulk-to-boundary expansion of bulk correlation functions does not converge for arbitrary operator configurations, and the resulting infinite sum need not commute with integration. Here we provide a heuristic argument explaining why these manipulations should nevertheless be valid in our setup, at least in generic situations, even though a fully rigorous justification is still lacking.

Let us first clarify when the various expansions are expected to converge. For the bulk-to-boundary expansion, the convergence condition is that one can choose a hemisphere $HS^{d}$ centered on the boundary ($z=0$) such that it contains the bulk operator while excluding all other operator insertions and defects. For the bulk-to-defect or boundary-to-defect expansions, the convergence condition is that one can choose a hemisphere centered on the defect which contains the operator being expanded, while excluding all other insertions.

With these convergence criteria in mind, let us now revisit and refine the previous argument.

We assume (A2'), which in particular states that the bulk-to-boundary expansion remains valid and unchanged in the presence of the defect. When $y \neq 0$, we can take the radial coordinate $z$ to be sufficiently small so that the bulk-to-boundary expansion of the stress tensor $T_{zi}(z,x,y)$ and the current $J^a_{z}(z,x,y)$ converges at the level of correlation functions:
\begin{equation}\label{TJ:bulktoboundary}
	\begin{split}
		T_{zi}(z,x,y) &= \sum_{V} b^{T}_{V}\, z^{\Delta_V-2} V(x,y)\,, \\
		J^a_{z}(z,x,y) &= \sum_{W} b^{J}_{W}\, z^{\Delta_W-1} W(x,y)\,,
	\end{split}
\end{equation}
where the sums run over all boundary operators, including both primaries and descendants.

The operators appearing in \eqref{TJ:bulktoboundary} are constrained by the Ward identities:
\begin{equation}\label{TJ:wardid}
	\begin{split}
		\partial_z\!\left(\frac{1}{z^{d-1}} T_{zi}\right) &= -\frac{1}{z^{d-1}} \partial_\mu T_{\mu i}\,, \\
		\partial_z\!\left(\frac{1}{z^{d-1}} J^a_{z}\right) &= -\frac{1}{z^{d-1}} \partial_\mu J^a_{\mu}\,,
	\end{split}
\end{equation}
where the index $\mu$ runs over all boundary directions, $\mu = 1,2,\ldots,d$. Plugging \eqref{TJ:bulktoboundary} into \eqref{TJ:wardid} and matching powers of $z$, we conclude that $V$ and $W$ must be total derivatives in $x$ and $y$. The only exceptions are the special cases when $\Delta_V = d+1$ and $\Delta_W = d$. Since these special cases are non-generic for QFT in AdS, we will not consider them in this work.

We now evaluate the integral by splitting it into two regions:
\begin{equation}
	\int d^{p}x\, d^{d-p}y
	=
	\int_{|y|<a} d^{p}x\, d^{d-p}y
	+
	\int_{|y|\geqslant a} d^{p}x\, d^{d-p}y \,,
\end{equation}
with a fixed cutoff $a$. We choose $a$ sufficiently small so that, for sufficiently small $z$, the bulk-to-defect expansion is valid in the region $|y|<a$. 

For the region $|y|<a$, the argument is almost the same as the one in sections \ref{subsec:displacement} and \ref{subsec:tilt}. In the limit $z\rightarrow0$, this part of the integral contributes the integrated displacement and tilt operators, plus boundary terms which depend on $a$.

For the region $|y|\geqslant a$, we first use the bulk-to-boundary expansion \eqref{TJ:bulktoboundary}. Since the operators are total derivatives, the integral reduces to a boundary term localized at $|y|=a$. At this locus, we can reliably use the boundary-to-defect expansion for the remaining integral, including the $x$-integral and the angular integral of $y$.

Therefore, the full integral effectively localizes in the vicinity of the defect and reduces to integrals of defect operators. The arguments in sections \ref{subsec:displacement} and \ref{subsec:tilt} then go through unchanged.\footnote{One may worry that \eqref{fTOint_scaling} could fail due to the cutoff $a$. This is not an issue: after summing the contributions from the regions $|y|<a$ and $|y|\geqslant a$, the final result is independent of $a$. Dimensional analysis therefore remains valid.}

\subsection{Shift symmetries in AdS}\label{app:shift}

We have argued that SSB of internal global symmetries leads to tilt operators with protected dimensions. In the case $p=d$, i.e., the defect is the whole boundary, the conclusion is $\Delta_\mathbb{T}=d$.

It is well-known that the scalar theory in AdS may have shift symmetries \cite{Bonifacio:2018zex,Blauvelt:2022wwa}:
\begin{equation}
	\begin{split}
		\phi(X)\rightarrow\phi(X)+C_{A_1\ldots A_k}X^{A_1}\ldots X^{A_k}\,,
	\end{split}
\end{equation}
where $C$ is a constant tensor and $X\in\mathbb{R}^{1,d+1}$ is the embedding coordinates of AdS$_{d+1}$ \cite{Costa:2014kfa}. The SSB of such shift symmetries leads to the protected dimension of the dual scalar field
\begin{equation}
	\begin{split}
		\Delta_k=d+k\,.
	\end{split}
\end{equation}
This can be verified explicitly in the free theory.\footnote{Here we do not consider the other conformal boundary condition which corresponds to $\Delta_k=-k$.} One may wonder what goes wrong for $k\geq1$, since $\Delta_k\neq d$. The answer is that our argument assumes that the global symmetries commute with the AdS isometries. For shift symmetries, this assumption is violated except for $k=0$.

This distinction is essential for studies of Wilson loops in gauge theories in AdS, which we will discuss in section \ref{subsec:YM_AdS}. The related effective string theory (EST) is a scalar theory in AdS$_2$ with shift symmetry of type $k=1$. Therefore, the prediction there is that the tilt operator, in the generalized sense, has dimension $\Delta_\mathbb{T}|_\text{EST}=2$, which matches the dimension of the displacement operators arising from the Wilson loop defect: $\Delta_{\mathbb{D}}|_W=2$. 

The shift symmetry in the flat space has a similar feature. In a scale-invariant theory, the shift symmetry $\mathcal{O}(x)\rightarrow\mathcal{O}(x)+c$ does not commute with the dilatation symmetry (unless $\Delta_{\mathcal{O}}=0$), resulting in the shift charge having non-zero but protected scaling dimension. This feature is essential for protection of the scaling dimension of a virial current, which is the obstruction to conformal invariance \cite{Gimenez-Grau:2023lpz}.

\section{Examples}\label{sec:example}
\subsection{Displacement operator in weakly coupled CFT}\label{section:weakly_coupled_CFT}

A broad class of defects can be constructed by inserting
\begin{equation}
	\begin{split}
		\mathcal{D}(y)=\exp\left(h\int d^px\,\mathcal{O}(x,y)\right),
	\end{split}
\end{equation}
where the integral is taken over the constant-$y$ slice. Correlators in the presence of the defect are defined by
\begin{equation}
	\begin{split}
		\braket{\mathcal{O}_1\ldots\mathcal{O}_n}_{\mathcal{D}}
		:=\frac{\braket{\mathcal{O}_1\ldots\mathcal{O}_n\mathcal{D}}}{\braket{\mathcal{D}}}\,.
	\end{split}
\end{equation}
With this definition, the translation symmetry along the transverse direction formally reads
\begin{equation}\label{pinningdefect_displacement}
	\begin{split}
		\left(\frac{\partial}{\partial y_1^i}
		+\frac{\partial}{\partial y_2^i}
		+\ldots+
		\frac{\partial}{\partial y_n^i}\right)
		\braket{\mathcal{O}_1\ldots\mathcal{O}_n}_{\mathcal{D}}
		+h\int d^px\,
		\braket{\partial_i\mathcal{O}(x,y)\mathcal{O}_1\ldots\mathcal{O}_n}_{\mathcal{D}}
		=0\,,
	\end{split}
\end{equation}
where $y_k^i$ denotes the $i$-th component of the transverse coordinate of $\mathcal{O}_k$. This suggests that the displacement operator is
\begin{equation}\label{displacement_classical}
	\begin{split}
		\mathbb{D}_i=h\partial_i\mathcal{O}\,.
	\end{split}
\end{equation}

Equation~\eqref{displacement_classical} should be understood as the classical expression for the displacement operator. In general, both the defect and the operators localized on it require renormalization. We refer to \cite{Popov:2025cha} for a detailed discussion of the renormalization of conformal defects. Suppose that
\[
\mathbb{D}_i=[h\partial_i\mathcal{O}]_{\rm ren}
\]
continues to define a local defect operator after renormalization, and that a renormalized version of \eqref{pinningdefect_displacement} is well defined. Then dimensional analysis implies $\Delta_{\mathbb{D}}=p+1$. However, to the best of the author's knowledge, a general justification of these assumptions is not presently available.

In what follows, we perform a one-loop check of the scaling dimension of the displacement operator in a class of weakly coupled defect CFTs.

Let us consider a CFT$_d$ satisfying the following assumptions:
\begin{itemize}
	\item there is a collection of scalar primary operators $\mathcal{O}_A$ with dimensions
	\[
	\Delta_A=p-a_A\epsilon\,,
	\]
	where $a_A=O(1)$ and $\epsilon$ is small;
	\item upon restriction to $p$ dimensions, there are no other CFT$_p$ scalar primary operators with transverse spin $0$ and dimensions near $\Delta=p$;
	\item apart from the transverse derivatives of the $\mathcal{O}_A$, there are no other CFT$_p$ primary operators with transverse spin $1$ and dimensions near $\Delta=p+1$.
\end{itemize}

Using these weakly relevant operators, we can generate a short RG flow supported on the $p$-dimensional subspace defining the defect:
\begin{equation}
	\begin{split}
		S=S_\text{CFT}
		+\sum_A\lambda^A\int d^p x\,\mathcal{O}_A(x,y=0)\,,
		\qquad
		x\in\mathbb{R}^p\,,\quad y\in\mathbb{R}^{d-p}\,.
	\end{split}
\end{equation}
The one-loop beta function is
\begin{equation}
	\begin{split}
		\beta^A(\lambda)
		=-a_A\epsilon\,\lambda^A
		+\frac{1}{2}S_{p-1}\sum_{B,C}C_{BC}^{\quad A}\lambda^B\lambda^C\,.
	\end{split}
\end{equation}
The putative IR fixed point is determined by
\begin{equation}
	\begin{split}
		\beta^A(\lambda_*)=0\,,
		\qquad A=1,2,\ldots,N\,,
	\end{split}
\end{equation}
where $N$ is the number of weakly relevant operators.

This procedure does not, by itself, guarantee the existence of an IR fixed point with $\lambda_*=O(\epsilon)$. In what follows, we assume that such a fixed point exists and examine its consequences.

We now ask whether the scaling dimension of the displacement operator is protected at order $O(\epsilon)$, namely whether $\Delta_{\mathbb{D}}=p+1$. To this end, we consider the one-loop scaling matrix
\begin{equation}
	\begin{split}
		\Gamma_{\alpha}^{\ \beta}(\lambda)
		=
		\Delta_\alpha^\text{UV}\delta_{\alpha}^{\ \beta}
		+S_{p-1}\sum_A C_{\alpha A}^{\quad \beta}\lambda^A\,.
	\end{split}
\end{equation}
Here $\alpha$ and $\beta$ label arbitrary CFT$_p$ primary operators. The matrix $\Gamma$ is block diagonal: operators in the same block have the same quantum numbers under rotations and internal symmetries, and only operators within the same block can mix. The eigenvalues of $\Gamma$ evaluated at $\lambda_*$ give the IR scaling dimensions at one loop.

To identify the displacement operator, we look for CFT$_p$ primaries with transverse spin $1$ and scaling dimensions $p+1+O(\epsilon)$. By assumption, the only such candidates are the transverse derivatives of the weakly relevant operators, $\partial_\perp\mathcal{O}_A$.

Let us therefore compute the entries of the scaling matrix in the subspace spanned by $\partial_\perp\mathcal{O}_A$. Since the $\mathcal{O}_A$ are assumed to be scalar primaries of the UV CFT$_d$, their two- and three-point functions are fixed by conformal symmetry:
\begin{equation}
	\begin{split}
		\braket{\mathcal{O}_A(x)\mathcal{O}_B(y)}
		&=
		\frac{\delta_{AB}}{\abs{x-y}^{2\Delta_A}}\,,
		\\
		\braket{\mathcal{O}_A(x)\mathcal{O}_B(y)\mathcal{O}_C(z)}
		&=
		\frac{C_{ABC}}
		{
			\abs{x-y}^{\Delta_A+\Delta_B-\Delta_C}
			\abs{x-z}^{\Delta_A+\Delta_C-\Delta_B}
			\abs{y-z}^{\Delta_B+\Delta_C-\Delta_A}
		}\,.
	\end{split}
\end{equation}
Taking transverse derivatives and then restricting all insertion points to the defect gives
\begin{equation}
	\begin{split}
		\braket{\partial_i\mathcal{O}_A(x)\partial_j\mathcal{O}_B(y)}
		&=
		\frac{2\Delta_A\delta_{AB}\delta_{ij}}
		{\abs{x-y}^{2\Delta_A+2}}\,,
		\\
		\braket{\partial_i\mathcal{O}_A(x)\mathcal{O}_B(y)\partial_j\mathcal{O}_C(z)}
		&=
		\frac{
			(\Delta_A+\Delta_C-\Delta_B)C_{ABC}\delta_{ij}
		}
		{
			\abs{x-y}^{\Delta_A+\Delta_B-\Delta_C}
			\abs{x-z}^{\Delta_A+\Delta_C-\Delta_B+2}
			\abs{y-z}^{\Delta_B+\Delta_C-\Delta_A}
		}\,.
	\end{split}
\end{equation}
It follows that
\begin{equation}
	\begin{split}
		C_{\partial_i\mathcal{O}_A\,\mathcal{O}_B}^{\qquad\partial_j\mathcal{O}_C}
		=
		\frac{\Delta_A+\Delta_C-\Delta_B}{2\Delta_C}
		C_{AB}^{\quad C}\delta_i^{\ j}
		=
		\frac{1}{2}C_{AB}^{\quad C}\delta_i^{\ j}
		+O(\epsilon)\,.
	\end{split}
\end{equation}
Substituting this into the one-loop scaling matrix, restricted to the $\partial_\perp\mathcal{O}_A$ sector, gives
\begin{equation}
	\begin{split}
		\Gamma_{\partial_i \mathcal{O}_A}^{\quad \partial_j\mathcal{O}_B}(\lambda)
		=
		\delta_i^{\ j}
		\left[
		(p+1-a_A\epsilon)\delta_A^{\ B}
		+\frac{1}{2}S_{p-1}\sum_C C_{AC}^{\ \ B}\lambda^C
		\right]\,.
	\end{split}
\end{equation}
Evaluating the matrix on the fixed-point coupling vector, we find
\begin{equation}
	\begin{split}
		\sum_A
		\Gamma_{\partial_i \mathcal{O}_A}^{\quad \partial_j\mathcal{O}_B}(\lambda_*)\lambda_*^A
		&=
		\delta_i^{\ j}
		\left[
		(p+1)\lambda_*^B+\beta^B(\lambda_*)
		\right]=
		(p+1)\lambda_*^B\delta_i^{\ j}\,.
	\end{split}
\end{equation}
Thus, to order $O(\epsilon)$, the operator
\begin{equation}
	\begin{split}
		\mathbb{D}_i
		=
		\sum_A\lambda_*^A\partial_i\mathcal{O}_A
	\end{split}
\end{equation}
has scaling dimension $p+1$. This is precisely the expected displacement operator dimension and is consistent with the general intuitive discussion above.

Importantly, this argument does not rely on any additional dynamical input beyond the assumed absence of extra CFT$_p$ primaries near dimensions $p$ in the scalar sector and $p+1$ in the transverse-spin-one sector. At one loop, the protection of the displacement operator follows purely from conformal symmetry and the fixed-point equation. It would be interesting to understand whether, in a general setting, the relation $\Delta_{\mathbb{D}}=p+1$ can be made manifest to all orders in conformal perturbation theory.

\subsection{Landau--Ginzburg models}

Consider a class of non-local Landau--Ginzburg models defined by
\begin{equation}\label{def:LGCFT}
	S =\mathcal{N}\int d^dx\, \phi(-\partial^2)^{\tfrac{d}{2}-\Delta_\phi}\phi+\sum_{n<\tfrac{d}{\Delta_\phi}} \frac{1}{n!}\lambda_n\,\mu^{d-n\Delta_\phi} \int d^d x\, \mathcal \phi^n(x)\,. 
\end{equation}
When all $\lambda_n=0$, the theory reduces to the generalized free theory (GFF), where $\phi$ is a generalized free field. The normalization factor $\mathcal{N}$ is chosen such that
\begin{equation}
	\begin{split}
		\braket{\phi(x)\phi(y)}_0=\frac{1}{\abs{x-y}^{2\Delta_\phi}}\,.
	\end{split}
\end{equation}
Turning on relevant interactions generates an RG flow, which may flow to an interacting long-range CFT in the IR.

As long as the bulk theory flows to an IR CFT, we can further introduce a $p$-dimensional localized deformation,
\begin{equation}
	\begin{split}
		S_{\text{defect}}=\sum_i \lambda^i \int d^p x\, \mathcal{O}_i(x,y=0)\,,
	\end{split}
\end{equation}
which induces an RG flow to a conformal defect. This construction encompasses a wide range of examples, including the long-range Ising, $O(N)$ (where $\phi$ is a $N$-dimensional vector), and Lee--Yang models \cite{Fisher:1972zz,Sak:1973oqx,Sak1977,Paulos:2015jfa,Behan:2017emf,Giombi:2019enr,Giombi:2020rmc,Giombi:2022gjj,Behan:2025ydd,Eustachon:2026vjn,Ghosh:2026gku}, which may admit various defect deformations \cite{Bianchi:2024eqm,Ge:2025fsm}.

\subsubsection{CFT viewed as boundary of AdS}
It is well known that a GFF can be realized as a conformal boundary condition of a massive free scalar field in AdS$_{d+1}$, with
\begin{equation}
	\begin{split}
		(mR_\text{AdS})^2=\Delta_\phi(\Delta_\phi-d)\,,
	\end{split}
\end{equation}
subject to the Breitenlohner--Freedman bound \cite{Breitenlohner:1982bm,Breitenlohner:1982jf}
\begin{equation}
	\begin{split}
		(mR_\text{AdS})^2\geqslant-\frac{d^2}{4}\,.
	\end{split}
\end{equation}
For these long-range CFTs, since the AdS bulk is not renormalized, they can be viewed as conformal boundary conditions of free scalar theories in AdS. Therefore, the general argument presented in this paper applies: the displacement and tilt operators associated with such defects have protected scaling dimensions.

\subsubsection{Caffarelli-Silvestre extension}
For this class of long-range CFTs, there is also an alternative protection mechanism based on the Caffarelli--Silvestre (CS) extension \cite{caffarelli2007extension}. In the CS picture, the non-local CFT can be embedded into a ($2\Delta_\phi+2$)-dimensional free scalar theory, which is a local and conformal but in fractional dimension \cite{Rajabpour:2011qr}. There, the non-local CFT appears as a conformal defect \cite{Paulos:2015jfa}. The displacement and tilt operators are then associated with the conservation of a local stress tensor and conserved currents in the extended space, which ensures their protection \cite{Ge:2025fsm}.

Both AdS and CS mechanisms imply that the anomalous dimensions of the displacement and tilt operators vanish to all orders in perturbation theory.

\subsubsection{$O(N)$ free theory}

The simplest example containing both displacement and tilt operators is perhaps given by $N$ copies of a GFF with $\Delta<p/2$. We introduce a $p$-dimensional thermal defect for the last copy,
\begin{equation}
	\begin{split}
		\lambda \int d^p x \,\phi_N^2\, .
	\end{split}
\end{equation}
For $\lambda<0$, the RG flow is indefinite, reflecting the fact that the $\phi_N^2$ potential is unbounded from below. For $\lambda>0$, the system flows to an IR fixed point $\lambda_*$. From the AdS point of view, this RG flow is known as the flow from Neumann boundary conditions in the UV to Dirichlet boundary conditions in the IR, giving
\begin{equation}
	\begin{split}
		\Delta_N^\text{IR}=p-\Delta\, .
	\end{split}
\end{equation}
The operator spectra in the other $N-1$ copies remain unchanged. Up to constant factors, the displacement operator is $\partial_\perp \phi_N^2$, which acquires anomalous dimension $p-2\Delta$ and thus has dimension $p+1$ in the IR, while the tilt operators are $\phi_i\phi_N$ with $i=1,\ldots,N-1$, and have dimension $p$.

For an extensive discussion of conformal defects in the interacting long-range $O(N)$ model, see \cite{Bianchi:2024eqm}.

\subsubsection{Comments on weakly coupled theories}

Finally, we note that in certain weakly coupled regimes, the RG flows described above can be analyzed explicitly at leading order. Some examples of short RG flows of this type fall into the category of section \ref{section:weakly_coupled_CFT}. For example, the $d$-dimensional long-range Ising model ($\phi^4$) with $d/2$-dimensional thermal defect ($\phi^2$) \cite{Ge:2025fsm}.

An important exception is provided by weakly relevant deformations of the form $\phi^{2n+1}$, such as the long-range Lee--Yang model realized as a conformal defect. In this case, the beta function does not contain a quadratic term in the coupling, due to the $\mathbb{Z}_2$ symmetry of the UV theory:
\begin{equation}
	\begin{split}
		\beta(\lambda)=(\Delta-p)\lambda+a\lambda^3+\ldots\,.
	\end{split}
\end{equation}
One may still find a short RG flow, but the perturbative analysis differs qualitatively from the cases considered in section \ref{section:weakly_coupled_CFT}. For example, $\lambda_*=O(\sqrt{\epsilon})$ instead of $O(\epsilon)$. We refer the reader to \cite{Ghosh:2026gku,Eustachon:2026vjn} for a detailed discussion of the long-range Lee--Yang model.

Another interesting exception is the magnetic defect in the interacting $O(N)$ model, for example
\begin{equation}
	\begin{split}
		\lambda \int d^d x\, \left(\sum_{i=1}^{N}\phi_i^2\right)^2
		+ g \int d^p x\, \phi_N \, .
	\end{split}
\end{equation} 
In this case, one can construct a short RG flow for $\lambda$ by tuning $\Delta_\phi$ to be slightly below $d/4$. One might then hope to obtain a short defect RG flow for $g$ by taking $p=d/4$. However, this leads to a contradiction: there is no candidate tilt operator, since the lightest candidate is $\phi_i\phi_N$, whose dimension is approximately $2p$. The resolution is that the defect RG flow cannot be short: $g_*$ must be at least $O(1)$. A similar analysis was carried out in the $O(N)$ Wilson--Fisher model in $d=4-\epsilon$ dimensions \cite{Cuomo:2021kfm}.

Nevertheless, the non-perturbative argument presented in this paper still applies, and the displacement and tilt operators remain protected.

\subsection{4D Maxwell theory}

Consider four-dimensional Maxwell theory:
\begin{equation}
	\begin{split}
		S[A]:=\frac{1}{4}\int d^3x\,dz\,\sqrt{g}\,F_{\mu\nu}F^{\mu\nu}\,.
	\end{split}
\end{equation}
By conformal invariance, Maxwell theory in AdS$_4$ is equivalent to the theory in the flat half-space
\begin{equation}
	\begin{split}
		ds^2=dz^2+d\mathbf{x}^2\,,\quad \mathbf{x}\in\mathbb{R}^3\,,\quad z\geqslant0\,.
	\end{split}
\end{equation}
We impose magnetic (Neumann) boundary conditions
\begin{equation}
	\begin{split}
		F_{zi}\big{|}_{z=0}=0\,,\quad i=1,2,3\,.
	\end{split}
\end{equation}

Under magnetic boundary conditions, gauge field degrees of freedom remain on the boundary, allowing us to define Wilson lines. We consider a straight Wilson line located at $x_1=x_2=0$:
\begin{equation}
	\begin{split}
		W=\exp\left(-iq\int_{\mathbb{R}}dx_3\, A_3\right)\,.
	\end{split}
\end{equation}
This defines a conformal line defect, with the displacement operator given by
\begin{equation}\label{Maxwell:displacement}
	\begin{split}
		\mathbb{D}_i(\tau)=-iqF_{i3}(0,0,0,\tau)\,,\quad i=1,2\,.
	\end{split}
\end{equation}
The two-point function of the displacement operator is
\begin{equation}
	\begin{split}
		\braket{\mathbb{D}_i(\tau_1)\mathbb{D}_j(\tau_2)}=\frac{2q^2}{\pi^2}\frac{\delta_{ij}}{(\tau_1-\tau_2)^4}\,.
	\end{split}
\end{equation}
In particular, we see that $\Delta_{\mathbb{D}}=2$.

We now consider how the displacement operator arises from the bulk-to-defect OPE of the stress tensor
\begin{equation}
	\begin{split}
		T_{\mu\nu}=F_{\mu\alpha}F_\nu^{\ \alpha}-\frac{1}{4}F_{\alpha\beta}F^{\alpha\beta}g_{\mu\nu}\,.
	\end{split}
\end{equation}
The relevant component is $T_{zi}$. We decompose it into the vacuum expectation value, single-photon, and double-photon contributions:
\begin{equation}
	\begin{split}
		T_{zi}&=\braket{T_{zi}}_W +\braket{F_{z\alpha}}_W :F_i^{\,\alpha}:+\braket{F_i^{\,\alpha}}_W :F_{z\alpha}: +:F_{z\alpha}::F_i^{\,\alpha}:\,.
	\end{split}
\end{equation}
Here we use normal ordering defined by $:\mathcal{O}:\equiv\mathcal{O}-\braket{\mathcal{O}}_W$. The non-vanishing expectation values of the field strength are
\begin{equation}
	\begin{split}
		\braket{F_{i3}}_W&=\frac{iq}{2\pi}\frac{x_i}{\rho^{3}}\,,\quad i=1,2\,, \\
		\braket{F_{z3}}_W&=\frac{iq}{2\pi}\frac{z}{\rho^{3}}\,,\quad \rho^2:=z^2+x_1^2+x_2^2\,.
	\end{split}
\end{equation}
All other components vanish. Therefore,
\begin{equation}
	\begin{split}
		T_{zi}&=-\frac{q^2 zx_i}{4\pi^2\rho^6}+\frac{iq}{2\pi}\frac{z}{\rho^{3}}:F_{i3}:+\frac{iq}{2\pi}\frac{x_i}{\rho^{3}}:F_{z3}:+:F_{z\alpha}::F_i^{\,\alpha}:\,.
	\end{split}
\end{equation}

We now examine the argument of section \ref{subsec:displacement} by integrating $T_{zi}$ over a constant-$z$ slice:
\begin{equation}
	\begin{split}
		\int d^3x\,T_{zi}(z,x)&=\frac{iq}{2\pi}\int d^3x\,\frac{z}{\rho^{3}}F_{i3}(z,x)+\frac{iq}{2\pi}\int d^3x\,\frac{x_i}{\rho^{3}}F_{z3}(z,x) \\
		&\quad+\int d^3x\,:F_{z\alpha}(z,x)::F_i^{\,\alpha}(z,x):\,.
	\end{split}
\end{equation}
By conservation, this expression is independent of $z$, so we may take the limit $z\rightarrow0$. In this limit, the first two integrals localize at $x_1=x_2=0$ due to the factor $z/\rho^3$, while the last term vanishes since $:F_{z\alpha}:=O(z)$ as a consequence of the magnetic boundary condition. Moreover, the second integral vanishes because the integrand is odd under $x_i\rightarrow -x_i$. We thus obtain
\begin{equation}
	\begin{split}
		\int d^3x\,T_{zi}(z,x)=\frac{iq}{2\pi}\lim_{z\rightarrow0^+}\int d^3x\,\frac{z}{\rho^{3}}F_{i3}(z,x)=iq\int_{\mathbb{R}} dx_3\,F_{i3}(0,0,0,x_3)\,.
	\end{split}
\end{equation}
Comparing with \eqref{Maxwell:displacement}, we find
\begin{equation}
	\begin{split}
		\int d^3x\,T_{zi}(z,x)=-\int_{\mathbb{R}} d\tau\, \mathbb{D}_i(\tau)\,.
	\end{split}
\end{equation}
This reproduces precisely the general relation derived in section~\ref{subsec:displacement}.

Finally, we emphasize that the above computation was performed entirely in flat space, made possible by the conformal invariance of four-dimensional Maxwell theory. In particular, no additional factors of $z$ arise from the metric or measure. In principle, the same computation can be carried out directly in AdS, yielding the same result. In contrast, for non-Abelian gauge theories in AdS$_4$, conformal invariance is broken at the quantum level, and such a mapping to flat half-space is no longer available.

\subsection{Yang--Mills in AdS}\label{subsec:YM_AdS}

We consider Yang--Mills theory in AdS$_{d+1}$, with $d=2,3$, subject to magnetic (Neumann) boundary conditions
\begin{equation}
	\begin{split}
		F_{iz}=0\,,
	\end{split}
\end{equation}
where $i=0,1,\ldots,d-1$ labels the boundary directions and $z$ is the radial coordinate. With these boundary conditions, the bulk gauge field is dual to a dynamical gauge field on the boundary \cite{Witten:2003ya,Marolf:2006nd}.

For $d=3$, one may instead impose electric (Dirichlet) boundary conditions. In that case, the bulk gauge field is dual to a conserved current in the boundary theory \cite{Witten:1998qj}. The current operators are then the only operators with scaling dimensions below $3$, and they can be used to construct topological surface defects. On such defects, a tilt operator exists, but there is no displacement operator, since the full conformal symmetry is preserved. It is believed that, as the AdS radius is increased from zero to infinity, the Dirichlet theory undergoes a phase transition and flows in the infrared to the Neumann theory \cite{Aharony:2012jf}.\footnote{\cite{Aharony:2012jf} proposed several possible scenarios for this phase transition, and it remains an open question which one is realized. See \cite{Ciccone:2024guw,DiPietro:2025ozw} for recent studies.} For the study of confinement in gauge theories, the magnetic boundary condition is the more natural choice.

With magnetic boundary conditions, we now insert an infinite Wilson line on the boundary. This defines a conformal line defect in the boundary theory, which is itself a non-local CFT. Unlike the Landau--Ginzburg examples discussed above, this setup involves a genuinely interacting bulk theory in AdS.

Our general argument then implies the existence of displacement operators associated with the breaking of transverse translations. When the AdS radius is much smaller than the confinement scale, the displacement operator can be identified with the field-strength component
\begin{equation}\label{D_vs_F}
	\begin{split}
		\mathbb{D}_i \sim i g_{\rm YM} R_{\rm AdS}^{3-d} F_{0i}\,,
	\end{split}
\end{equation}
where $0$ and $i$ label the directions parallel and transverse to the line defect. Thus, in $d=2$ there is one displacement operator, while in $d=3$ there are two. Equation~\eqref{D_vs_F} is exact in Maxwell theory. In non-Abelian gauge theory, it receives higher-order corrections, and the operator must be inserted inside the Wilson line with the appropriate trace over color indices.

In \cite{Gabai:2025hwf}, the existence of the displacement operator was assumed as a hypothesis and shown to be consistent with the predictions of effective string theory (EST). EST provides a universal long-distance description of confining flux tubes and interfaces in spin systems \cite{Luscher:1980ac,Polchinski:1991ax,Luscher:2004ib,Aharony:2011gb,Dubovsky:2012sh,Dubovsky:2013gi,Aharony:2013ipa,Dubovsky:2014fma,Hellerman:2014cba,Dubovsky:2015zey,Hellerman:2016hnf,Billo:2006zg,Billo:2016cpy,Athenodorou:2022pmz,Baffigo:2023rin,Lima:2025sqa,Lopes:2026erz}. In AdS, EST predicts that the transverse modes on the long string have scaling dimension $2$ and transverse spin $1$, precisely matching the quantum numbers of the displacement operator for a conformal line defect. The main result of this work gives a first-principles explanation of this structure.

\section{Conclusion and outlook}\label{sec:conclusion}

The argument presented in this work shows that the existence and protection of displacement and tilt operators do not rely on the presence of a boundary stress tensor or conserved currents in a non-local defect CFT defined through AdS. Instead, they follow from a Goldstone-type mechanism: the interplay between bulk locality in AdS and the spontaneous symmetry breaking induced by the boundary defect. In this sense, the protected defect operators can be understood as boundary manifestations of Goldstone modes. The protection of their operator dimensions is therefore not an intrinsic property of the boundary theory alone, but rather a consequence of its embedding into a local higher-dimensional system.

We expect the same conclusions to hold when a local QFT in AdS is coupled to an additional boundary theory, since the AdS bulk remains local. One example is the RG fixed point obtained by coupling a generalized free field to a short-range CFT \cite{Behan:2017emf,Behan:2025ydd,Eustachon:2026vjn}.

A direct consequence of this work is that conformal-defect techniques based on displacement and tilt operators continue to apply in this setting \cite{Drukker:2022pxk,Gabai:2025zcs,Girault:2025kzt,Kong:2025sbk,Belton:2025ief,Drukker:2025dfm}.

Our analysis highlights a broader lesson: even when the boundary theory is non-local, the existence of a local bulk description in AdS imposes strong constraints on the operator spectrum and couplings of the boundary CFT \cite{Bena:1999jv,Hamilton:2005ju,Hamilton:2006az,Kabat:2011rz,Levine:2023ywq,Meineri:2023mps,Levine:2024wqn,Loparco:2025aag}. In particular, bulk locality acts as an organizing principle that enforces universal features of defect dynamics.

The examples considered here illustrate that this mechanism applies across a broad range of settings, from weakly coupled constructions to strongly interacting gauge theories in AdS. In particular, the appearance of displacement operators for Wilson line defects provides a nontrivial consistency check of the general picture of confinement in AdS \cite{Callan:1989em,Aharony:2012jf,Ciccone:2024guw,Ciccone:2025dqx,Gabai:2025zcs,Gabai:2026myo}.

This work focused only on continuous internal $0$-form symmetries, which constitute a special class of generalized symmetries \cite{Gaiotto:2014kfa}. Our results provide an example in which spontaneous symmetry breaking in AdS gives rise to Goldstone modes with masses of order $O(1/R_{\rm AdS})$, rather than exactly massless excitations. In recent years, Goldstone mechanisms associated with generalized symmetries have been extensively studied, primarily in flat space \cite{Gaiotto:2014kfa,Lake:2018dqm,Hofman:2018lfz,Hidaka:2020ucc,Armas:2018zbe,Distler:2021qzc,Yuan:2019geh,Hirono:2022dci,GarciaEtxebarria:2022jky,Afxonidis:2023tup,Damia:2023gtc,Berean-Dutcher:2025ohp}. Extending these ideas to AdS would be an interesting direction for future work.\footnote{The spontaneous breaking of one-form symmetries in AdS was recently discussed in \cite{Ankur:2026ylr}.}

Several open questions remain. A more systematic understanding of the convergence issues discussed in section~\ref{app:improve} would clarify the precise scope of the argument and may lead to a rigorous proof. It would also be interesting to explore how these protected operators constrain correlation functions in non-local defect CFTs, for instance through the conformal bootstrap and defect CFT techniques. More broadly, our results emphasize the question of how far bulk locality can be used to classify and constrain non-local conformal theories. We expect that the interplay between symmetry, bulk locality, and defect dynamics will continue to provide a useful framework for uncovering universal structures beyond the standard setting of local CFTs.

\section*{Acknowledgement}
I thank Jonas Dujava, Barak Gabai, Victor Gorbenko, Ziwen Kong, Yu Nakayama, and Bendeguz Offertaler for useful discussions. I especially thank Gabriel Cuomo for the insightful discussion at the early stage of this project. The work of JQ was supported by World Premier
International Research Center Initiative (WPI), MEXT, Japan, and by the Center for Data-Driven Discovery, Kavli IPMU (WPI).

\bibliography{defect_LRCFT}
\bibliographystyle{utphys}
	
\end{document}